\documentclass[aps,prd,amsmath,amssymb,twocolumn,superscriptaddress,preprintnumbers]{revtex4-1}
\usepackage[utf8]{inputenc} 
\usepackage{graphicx}
\usepackage{dcolumn}
\usepackage{bm}
\usepackage[bookmarks, pagebackref=false]{hyperref}
\usepackage[usenames,dvipsnames]{xcolor}
\usepackage[normalem]{ulem}
\definecolor{rossoCP3}{cmyk}{0,.88,.77,.40}
\definecolor{blaa}{rgb}{0.2,0.2,0.6}
\hypersetup{colorlinks, 
	bookmarksopen, 
	bookmarksnumbered,
	citecolor=blaa, 		
	linkcolor=rossoCP3,	
	urlcolor=rossoCP3,			
}
\usepackage{multirow}

\usepackage{natbib}
\usepackage{slashed}
\newcolumntype{x}[1]{>{\centering\arraybackslash\hspace{0pt}}p{#1}}

\graphicspath{{./figs/}}

\begin{document}
 
\title{ \LARGE  \color{rossoCP3} Naturalness\\of\\ lepton non-universality and muon
 g-2}
\author{Giacomo {\sc Cacciapaglia}}
\thanks{{\scriptsize Email}: \href{mailto:g.cacciapaglia@ipnl.in2p3.fr}{g.cacciapaglia@ipnl.in2p3.fr}; {\scriptsize ORCID}: \href{https://orcid.org/0000-0002-3426-1618}{0000-0002-3426-1618}}
\affiliation{Institut de Physique des 2 Infinis de Lyon (IP2I), UMR5822, CNRS/IN2P3,  F-69622 Villeurbanne Cedex, France}
\affiliation{University of Lyon, Universit\'e Claude Bernard Lyon 1, F-69001 Lyon, France}

\author{Corentin {\sc Cot}}
 \thanks{{\scriptsize Email}: \href{mailto:cot@ipnl.in2p3.fr}{cot@ipnl.in2p3.fr}; {\scriptsize ORCID}: \href{https://orcid.org/0000-0001-5845-6500}{0000-0001-5845-6500}}
\affiliation{Institut de Physique des 2 Infinis de Lyon (IP2I), UMR5822, CNRS/IN2P3,  F-69622 Villeurbanne Cedex, France}
\affiliation{University of Lyon, Universit\'e Claude Bernard Lyon 1, F-69001 Lyon, France}

\author{Francesco {\sc Sannino}}
\thanks{{\scriptsize Email}: \href{mailto:sannino@cp3.sdu.dk}{sannino@cp3.sdu.dk}; {\scriptsize ORCID}: \href{https://orcid.org/0000-0003-2361-5326}{ 0000-0003-2361-5326}}
\affiliation{Scuola Superiore Meridionale, Largo S. Marcellino, 10, 80138 Napoli NA, Italy}
\affiliation{Danish Institute for Advanced Study, Univ. of Southern Denmark, Campusvej 55, 5230 Odense M, Denmark }
\affiliation{Dipartimento di Fisica, E. Pancini, Universit\'a di Napoli Federico II, INFN sezione di Napoli, Complesso Universitario di Monte S. Angelo Edificio 6, via Cintia, 80126 Napoli, Italy}
\affiliation{CP$^3$-Origins, Univ. of Southern Denmark, Campusvej 55, 5230 Odense M, Denmark} 

\begin{abstract}  
We show that the observed anomalies in the lepton sector can be explained in extensions of the Standard Model that are natural and, therefore, resolve the Higgs sector hierarchy problem.  The scale of new physics is around the TeV and Technicolor-like theories are ideal candidate models. 
\end{abstract}
 
\maketitle

Naturalness has been a driving principle for the past decades when modelling and searching for new physics. The observation of the Higgs boson, with properties similar to the ones predicted by the Standard Model (SM), not followed by the discovery of new particles has cast doubts on this guidance principle.  In truth, the natural models disfavoured by experiments are vanilla extensions that predict a sizeable number of states below the TeV scale, such as the constrained minimal supersymmetric SM and composite Higgs models with several light top partners. 

In this paper we show that the experimental observation of lepton non-universal processes together with the muon g-2 anomaly brings back naturalness as a prime motor for generating new models of fundamental interactions. Marrying Technicolor-like models \cite{Weinberg:1975gm,Dimopoulos:1979es} with SM fermion partial compositeness \cite{Kaplan:1991dc}, we show that it is possible to accommodate the observed anomalies. The composite scale is around $2$~TeV, which is the natural techni-fermion condensation scale. As for the Higgs nature, the fact is that strong dynamics is not easy to solve and the jury is still out on what this state could be.  For example, it could emerge as a near dilaton \cite{Bardeen:1985sm,Yamawaki:1985zg,Sannino:1999qe,Hong:2004td,Dietrich:2005jn,Appelquist:2010gy,Goldberger:2007zk} in walking Technicolor dynamics \cite{Holdom:1988gs,Cohen:1988sq}.  In this case, the associated effective action is implemented, following Coleman~\cite{coleman_1985}, by saturating the underlying trace anomaly of the theory. In recent years there has been interest in this type of effective framework~\cite{Hong:2004td,Dietrich:2005jn,Goldberger:2008zz,Appelquist:2010gy,Hashimoto:2010nw,Matsuzaki:2013eva,Golterman:2016lsd,Hansen:2016fri,Golterman:2018mfm} partially fueled by lattice investigations. These are about $SU(3)$ gauge theories with $N_f= 8$ fundamental Dirac fermions~\cite{Appelquist:2016viq,Appelquist:2018yqe,Aoki:2014oha,Aoki:2016wnc}, as well as $N_f = 3$ symmetric 2-index Dirac fermions (sextets)~\cite{Fodor:2012ty,Fodor:2017nlp,Fodor:2019vmw}. The latter are known as Minimal Walking Technicolor~\cite{Hong:2004td,Sannino:2004qp,Dietrich:2005jn,Evans:2005pu} models. For all these models, the lattice collaborations reported evidence for the existence of a light singlet scalar particle. 
A light Higgs state could also emerge due to top quark corrections, as pointed out in \cite{Foadi:2012bb}, and/or emerging from non QCD-like dynamics \cite{Hong:2004td,Foadi:2012bb}.   Another possibility is the composite Goldstone Higgs paradigm \cite{Kaplan:1983fs,Contino:2003ve} that is, however, disfavoured by the muon $g-2$ anomaly, as we will show. For a general review of composite dynamics, we refer the reader to \cite{Cacciapaglia:2020kgq}.
A discussion of naturalness and the muon $g-2$ anomaly in supersymmetric models can be found in \cite{LI2018255,Baer:2021aax}.

The Fermilab collaboration recently presented their first preliminary  measurement of the muon anomalous magnetic moment \cite{Abi:2021gix}, confirming a discrepancy  from the SM value (by $3.3\sigma$).  Combining this result with the BNL E821 one \cite{Bennett:2006fi} leads to the experimental average of 
\begin{equation}
a_{\mu} ({\mathrm{Exp}}) = 116 \, 592 \, 061 (41) \times 10^{-11} \quad (0.35\,{\mathrm{ppm}}) \ ,
\end{equation} 
 with an overall deviation from the SM central value \cite{Aoyama:2020ynm} of
  \begin{equation}
\Delta a_{\mu} = a_{\mu} ({\mathrm{Exp}})  - a_{\mu} ({\mathrm{SM}}) = (251 \pm 59) \times 10^{-11} \ ,
\end{equation} 
corresponding to a significance of $4.2\sigma$ \cite{Abi:2021gix}. 

Few weeks earlier the LHCb~\cite{Aaij:2021vac} collaboration presented their updated measurement of the ratio $R_K$, which is an important test of  lepton universality in the SM:
\begin{equation} \label{eq:RKexp}
R_K =\frac{{\rm BR} \left ( B^+ \to K^+ \mu ^+ \mu ^-\right )}{{\rm BR} \left ( B^+ \to K^+ e^+ e^-\right )}= 0.846^{+0.044}_{-0.041} \, .
\end{equation}
This measure also deviates from the SM predictions at a $3.1\sigma$ level. 

By comparing elementary and composite extensions of the SM, we show that Technicolor-like models \cite{Weinberg:1975gm,Dimopoulos:1979es} yield a natural interpretation of these results with imminent consequences for collider physics. 
We start by noticing that both $R_K$ and $\Delta a_\mu$ are observables strongly related to the Yukawa sector of the SM, meaning that to explain the anomalies the new physics is related to the fermion mass generation mechanism. In fact, to account for the observed $\Delta a_\mu$ anomaly a new Higgs sector is crucial. 
 
To better elucidate our point we employ a model that, depending on the underlying dynamics, can be interpreted as either elementary, or partially elementary, or fully composite. Without further ado, we add to the SM Lagrangian the following interactions: 
\begin{multline}  \label{eq:Yuk2}
- \mathcal{L}_{\rm NP} =
y_L^{ij} \, L^i  \mathcal{F}_{L}  (\mathcal{S}_{E}^j)^\ast+
y_E^{ij} \, (E^i)^c \mathcal{F}^c_{N} \mathcal{S}_{E}^j +\\
y_Q^{ij} \, Q^i \mathcal{F}_{L} (\mathcal{S}_{D}^j)^\ast+
y_U^{ij} \, (U^i)^c \mathcal{F}_{E}^c \mathcal{S}_{D}^j+
y_D^{ij} \, (D^i)^c \mathcal{F}_{N}^c \mathcal{S}_{D}^j+ \\
\sqrt{2} \kappa (\mathcal{F}_L \mathcal{F}_N^c + \mathcal{F}_E \mathcal{F}_L^c) \phi_H +
\hbox{h.c.} 
\end{multline}
where $\cal F$ and $\cal S$ are (Weyl) fermions and scalars, respectively, with multiplicity $N_{\rm TC}$. We will interpret this multiplicity either as a global number or as the dimension of the fundamental representation of a new confining gauge group $\mathcal{G}_{\rm TC}$. Their specific quantum numbers are listed in Table~\ref{tab:1}. In the Lagrangian \eqref{eq:Yuk2}, $L$, $E$ and $Q$ are the usual chiral SM fields, while $\phi_H$ is a state with the Higgs doublet quantum numbers. 

We list below the various underlying interpretations of the model: 
\begin{itemize}
\item[i)]{New elementary fermions and scalars with renormalisable interactions that radiatively contribute to the SM fermion masses. An interesting limit occurs when the flavour structure of the SM fermion masses is  radiatively generated by loops of $\cal F$  and $\cal S$, hence the flavour structure is encoded in the new Yukawa couplings rather than in the Higgs couplings. Here, $\cal F$ are vector-like (thus they have both chiralities) and have a tree-level mass $M_{\cal F}$.}
\item[ii)]{The Higgs of the SM is replaced by a Technicolor composite state of (non) Goldstone nature, while the new scalars are still elementary, as put forward in fundamental partial compositeness \cite{Sannino:2016sfx}. 
This implies that $\kappa$ models the coupling between the  effective Higgs field and its constituents. The fermions $\cal F$ can be chiral or vector-like, depending on the details of the model.}
\item[iii)]{As in ii) but with the new scalars interpreted as effective operators  made of new extended Technicolor fermions \cite{Cacciapaglia:2020kgq}. This limit covers also models of partial compositeness presented in \cite{Barnard:2013zea,Ferretti:2013kya}.}
\end{itemize}
For the cases ii) and iii), $\mathcal{G}_{\rm TC}$ confines at low energies and generates the Higgs as a composite (Goldstone) meson. Here, in addition to their eventual bare masses, the new fermions $\cal F$ and scalars $\cal S$ will acquire a dynamical mass upon condensation, equal to a fraction of the condensation scale $\Lambda_{\rm TC}$.

\begin{table}
\begin{tabular}{l|c|c|c|c|}
  & $\mathcal{G}_{\rm TC}$ & $SU(3)_c$ & $SU(2)_L$ & $U(1)_Y$ \\\hline
$\mathcal{F}_{L}$ & $\bf F$ & $1$ & $2$ & $Y$ \\ 
$\mathcal{F}_{N}^c$ &  $\bf \bar F$ & $1$ & $1$ & $-Y-1/2$ \\ 
$\mathcal{F}_{E}^c$ &  $\bf \bar F$ & $1$ & $1$ & $-Y+1/2$ \\  \hline
$\mathcal{S}_{E}^j$ & $\bf F$ & $1$ & $1$ & $Y-1/2$ \\
$\mathcal{S}_{D}^j$ & $\bf F$ & $3$ & $1$ & $Y+1/6$  \\\hline
\end{tabular}
\caption{\em\label{tab:1} Quantum numbers of the new fermions $\cal F$ and scalars $\cal S$ in the model. $\mathcal{G}_{\rm TC}$ can be considered either gauged, as in composite scenarios, or global in a renormalisable model of flavour.} 
\end{table}

In Table~\ref{tab:cCH} we summarise the expressions that we use to analyse the data relevant for the flavour and muon $g-2$ anomalies. In particular, $c_{b_L \mu_L}$ dominates the leptonic $B$ decays relevant for the $R_K$ and $R_{K^\ast}$ measurements, while $C_{B\bar B}$ encodes the strongest constraint from the $B_s$ mixing \cite{DAmico:2017mtc}. The latter constraint provides an upper bound on the quark Yukawa combination $(y_Q y_Q^\ast)_{bs}$, which we fix to suitable values in our numerical analysis. In the second column of the Table we show the perturbative estimate of the muon $g-2$ stemming from loops of the heavy fermions and scalars.  The associated Naive Dimensional Analysis (NDA) estimate from composite dynamics is displayed in the third column.  For $\Delta a_\mu$, the relevant charges read $q_{\cal F} = Y+1/2$ and $q_{{\cal S}_E} = Y-1/2$ with $Y$ the hypercharge of the fermion doublet ${\cal F}_L$. For the elementary case the hypercharge is chosen to yield integer charges for the non QCD-coloured states, in particular for the plots we assume $Y=1/2$. For the composite case the hypercharge depends on the details of the model, in particular on $N_{\rm TC}$. 
For example, for $N_{\rm TC}=2$ we have $\mathcal{G}_{\rm TC}=SU(2)$ with $Y=0$, leading to the model studied in \cite{Ryttov:2008xe,Galloway:2010bp}. In this model with two Dirac techni-fermions, one can simultaneously accommodate composite Goldstone Higgs and Technicolor models \cite{Cacciapaglia:2014uja} and lattice have confirmed the pattern of chiral symmetry breaking $SU(4)$ to $Sp(4)$ \cite{Lewis:2011zb,Hietanen:2013fya,Arthur:2016dir}.  For the traditional Technicolor case of $N_{\rm TC}=3$ it is possible to assume the value  $Y=1/2$. Additionally, in the composite theory, $g_{\rm TC} \sim \frac{4\pi}{\sqrt{N_{\rm TC}}}$ takes care of varying number of Technicolors and the Yukawa couplings in Eq.~\eqref{eq:Yuk2} appear in the combinations $\frac{y y'}{g_{\rm TC}}$, which is assumed to be a perturbative coupling in the effective field theory description of the composite models. 
In the NDA estimates, it is not possible to know, a priori, the sign of the coefficients. Hence, we take the sign from the loop effects with equal masses: for $\Delta a_\mu$ the contribution is positive, as required by the measured muon $g-2$, for $Y = 0$ or positive.

\begin{table*}
$$ \begin{array}{c|c|c}
\hbox{Coefficient} & \hbox{Perturbative one-loop result} & \hbox{Non-perturbative NDA}\\ \hline
c_{b_L \mu_L}
&  \displaystyle N_{\rm TC}  \frac{(y_{L} y_L^\dagger)_{\mu\mu} (y_{Q} y_{Q}^\dagger)_{bs}}{(4\pi)^2 M_{{\cal F}}^2} \frac{1}{4} F(x,y)
&  \displaystyle \frac{(y_{L} y_L^\dagger)_{\mu\mu} (y_{Q} y_{Q}^\dagger)_{bs}}{g_{\rm TC}^2 \Lambda_{\rm TC}^2}\\[5mm]
C_{B\bar B}
&  \displaystyle N_{\rm TC}  \frac{ (y_{Q} y_{Q}^\dagger)_{bs}^2}{(4\pi)^2 M_{{\cal F}}^2} \frac{1}{8} F(x,x)
& \displaystyle  \frac{ (y_{Q} y_{Q}^\dagger)_{bs}^2}{g_{\rm TC}^2 \Lambda_{\rm TC}^2}\\[3mm]
\multirow{2}{*}{$\Delta a_\mu$}
& \displaystyle  N_{\rm TC} \frac{m_\mu (y_{L} y_E^\dagger)_{\mu\mu} \kappa v_{\rm SM}}{(4\pi)^2 M_{{\cal F}}^2 }\left[2 q_{{\cal S}_E} F_{LR} (y) + 2 q_{{\cal F}}  G_{LR}(y)\right] +
& \multirow{2}{*}{$\displaystyle   \frac{m_\mu^2}{ \Lambda_{\rm TC}^2 } \left( 1 + \frac{(y_{L} y_L^\dagger)_{\mu\mu}}{{g_{\rm TC}^2}}\right)$}\\[4mm]
& \displaystyle    N_{\rm TC} \frac{m_\mu^2 (y_{L} y_L^\dagger)_{\mu\mu}}{(4\pi)^2 M_{{\cal F}}^2 }\left[2 q_{{\cal S}_E}F_7(y) + 2 q_{\cal F} \tilde{F}_7(y)\right]  &  \\
\end{array}
$$
\caption{\em\label{tab:cCH}  Coefficients of the low-energy operators impacting  leptonic $B$ decays ($c_{b_L \mu_L}$), $B_s$ mixing ($C_{B\bar B}$) and the muon $g-2$, estimated via perturbative loops of the heavy fermions and scalars (second column) \cite{DAmico:2017mtc,Calibbi:2018rzv} or via their strongly-interacting NDA (third column).  For $\Delta a_\mu$, the relevant charges read $q_{\cal F} = Y+1/2$ and $q_{{\cal S}_E} = Y-1/2$. In the composite models, $m_\mu \sim N_{\rm TC} \frac{(y_{L} y_E^\dagger)_{\mu\mu} \kappa v_{\rm SM}}{(4 \pi)^2}$. The loop functions, where $x=M_{{\cal S}_D}^2/M_{\cal F}^2$ and $y=M_{{\cal S}_E}^2/M_{\cal F}^2$, are given in the appendix.}
\end{table*}

\begin{figure}
\begin{center}
\includegraphics[width=0.4\textwidth]{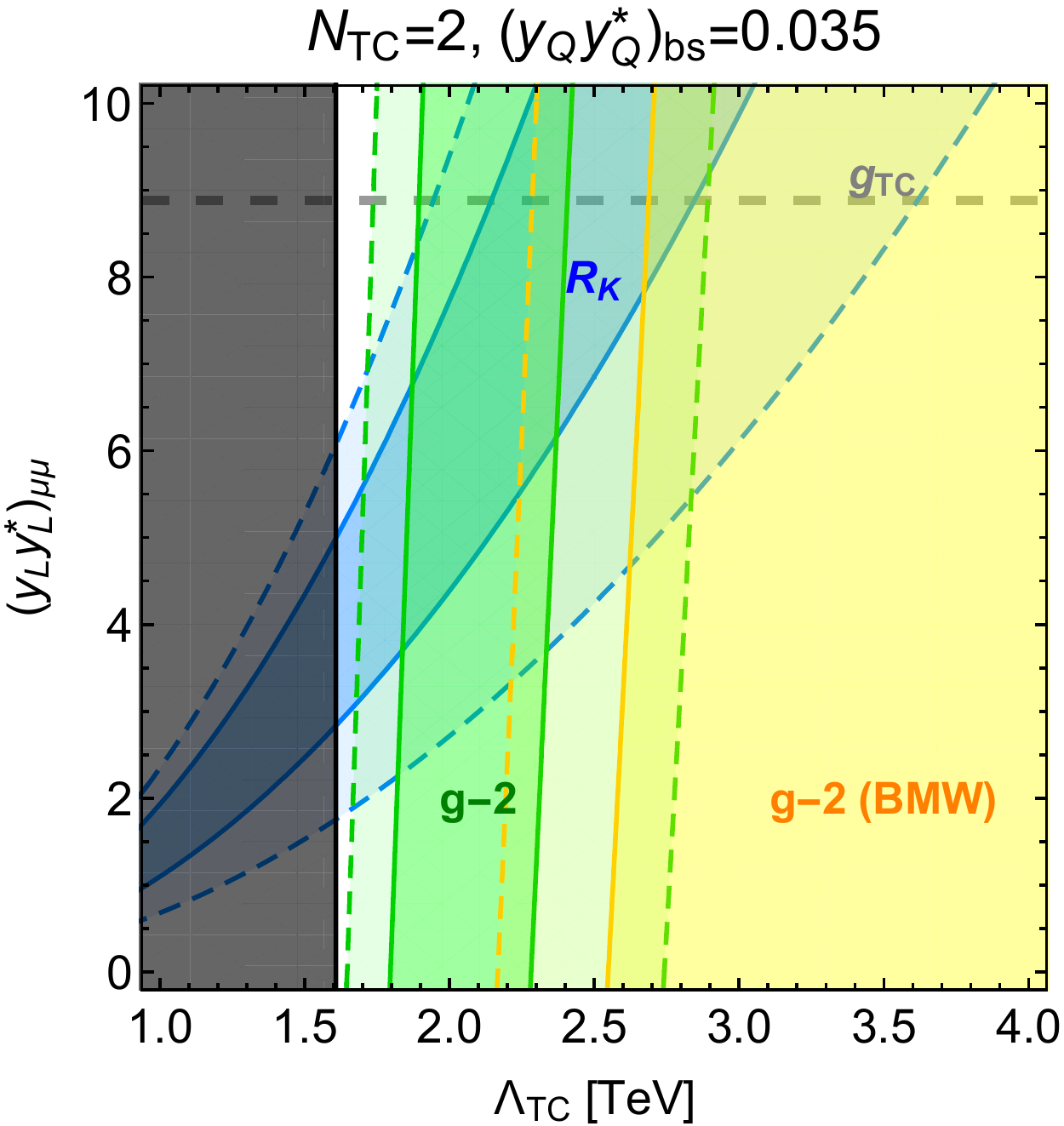}
\end{center}
\caption{ \label{fig:NDA}\em Parameter space of the model following the NDA estimates. The green bands fit the muon $g-2$ anomaly following the world average for the SM prediction \cite{Aoyama:2020ynm}. The blue regions fit $R_K$, while the shaded gray area is excluded by the $B_s$ oscillations. The solid (dashed) lines indicate $1\sigma$ ($2\sigma$) regions. The yellow area corresponds to the muon $g-2$ anomaly if the recent BMW lattice calculation \cite{Borsanyi:2020mff} is used for the hadronic photon polarisation. The horizontal grey line marks the value of the muon $y_L$ for which $(y_L y^\ast_L)_{\mu \mu} = g_{\rm TC}$.}
\end{figure}

We start with the new physics explanation coming from a composite nature of the SM Higgs sector together with  (fundamental) partial compositeness for the fermions. The results are summarised in Fig.~\ref{fig:NDA} for a model featuring $N_{\rm TC} = 2$ and $Y=0$. From the NDA estimate, $\Delta a_\mu$ is mostly sensitive to the scale of new physics $\Lambda_{\rm TC}$ with a mild dependence on the left-handed muon Yukawa, as it is visible from the near vertical green bands. They are based on the deviation from the SM world averages reported in \cite{Aoyama:2020ynm}.  Here we learn that the muon $g-2$ anomaly requires a rather low composite scale of the order of $2$~TeV, independently on other parameters of the model. At the same time, in order to address the $R_K$ anomaly, a large left-handed muon Yukawa coupling is required with values compatible with composite dynamics represented by the region below the horizontal gray dashed line.

The low composite scale disfavours composite Goldstone Higgs (CGH) dynamics, which needs a little hierarchy between the composite and electroweak scale, thus leaving a Technicolor-like explanation for the anomalies. This conclusion is also supported by independent estimates of $\Delta a_\mu$ for CGH dynamics  \cite{Frigerio:2018uwx} which read: 
 \begin{equation}
 \Delta a_\mu ({\rm CGH}) \approx \frac{g_\ast^2}{(4\pi)^2} \frac{m_\mu^2}{m_{\ast}^2}  = \frac{v^2_{\rm SM}}{f^2_{\rm CGH}}\frac{m_\mu^2}{\Lambda_{\rm TC}^2}\,,
\end{equation}
with $m_\ast = g_\ast  f_{\rm CGH}$ and $\Lambda_{\rm TC} \approx 4\pi v_{\rm SM} \approx 2$~TeV. Here, $f_{\rm CGH}$ is the Goldstone Higgs decay constant, and it is expected to assume a value substantially above the TeV scale to ensure a Goldstone nature of the Higgs and to reduce the tension with electroweak precision bounds. 
 
The situation can improve for the CGH scenario if the latest lattice results from the BMW collaboration \cite{Borsanyi:2020mff}, not included in the world average, are considered as the true contribution to the hadronic vacuum polarisation of the photon. These are represented by the yellow bands in Fig.~\ref{fig:NDA}. However, in the BMW scenario it is harder to account for $R_K$ for fixed $N_{\rm TC}$. This is so since it would require a too large muon left-handed Yukawa coupling. These results hold for both cases ii) and iii) on the model list. A way out is to increase $N_{\rm TC}$. 

\begin{figure*}
\begin{center}
\includegraphics[width=0.4\textwidth]{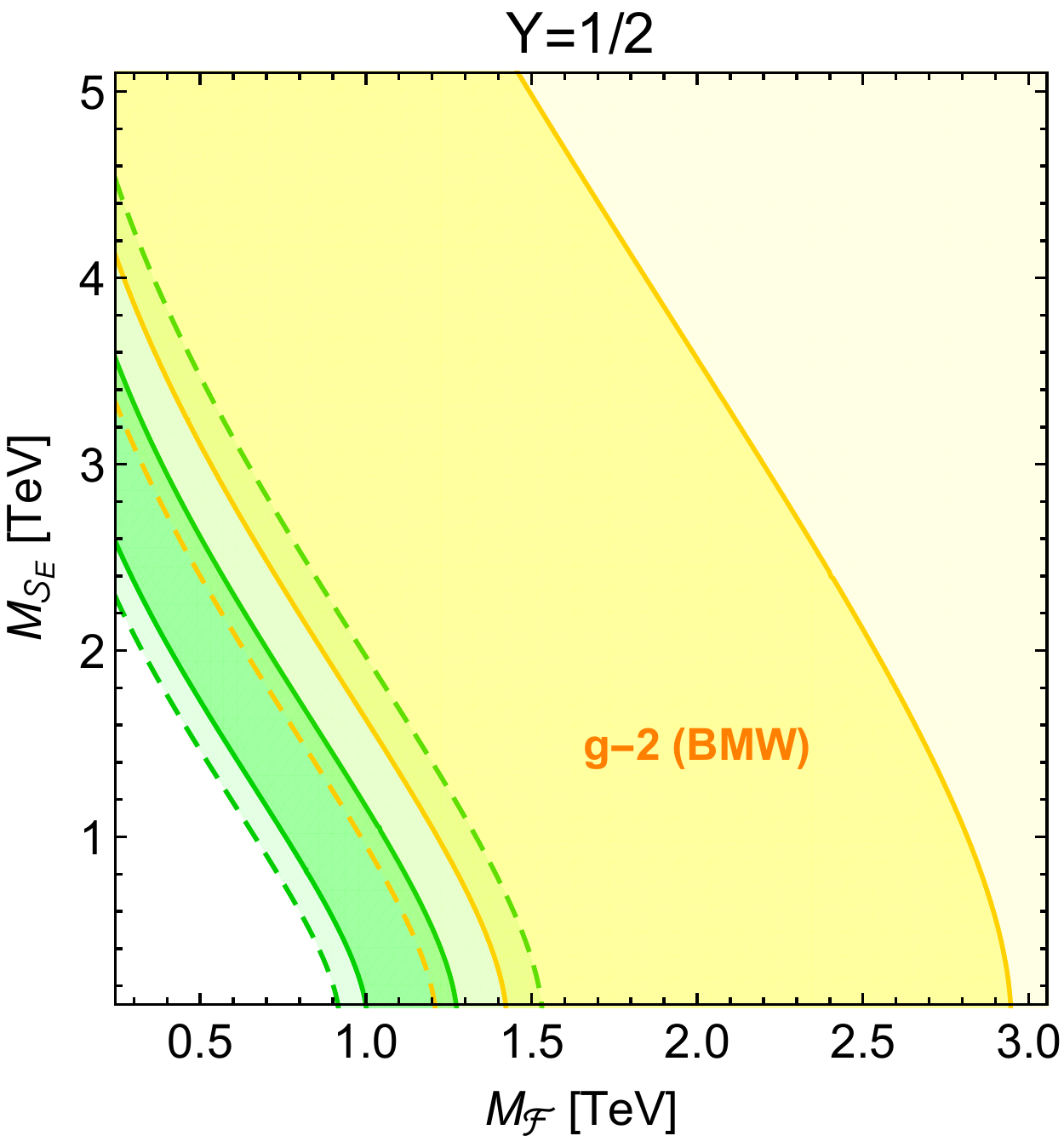} \hspace{0.7cm}
\includegraphics[width=0.4\textwidth]{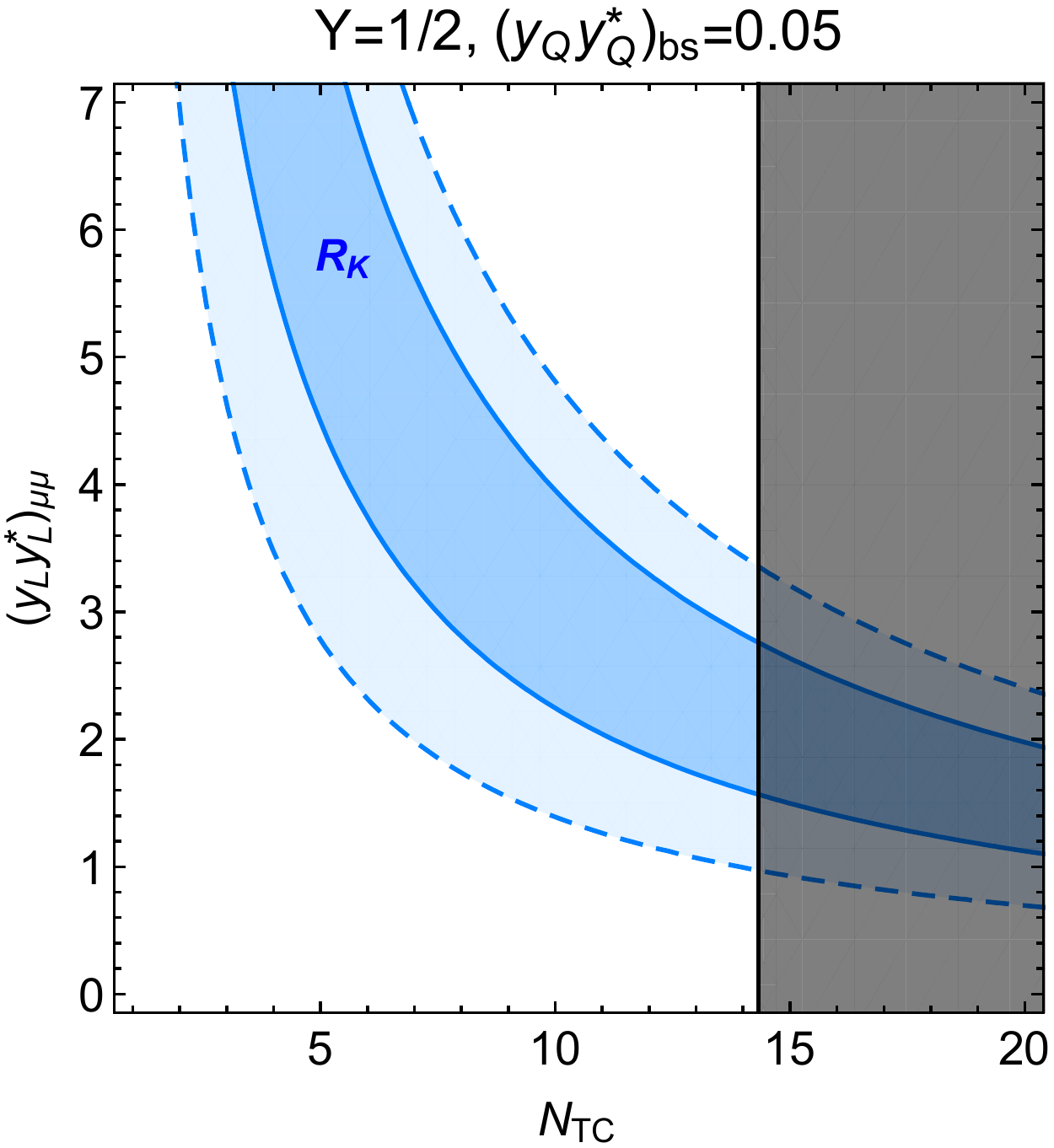}
\end{center}
\caption{\label{fig:loop}\em Parameter space of the model following the loop estimates from Table~\ref{tab:cCH}. In the left panel, we show the regions preferred by the muon $g-2$ anomaly as a function of the two relevant masses, $M_{\cal F}$ and $M_{{\cal S}_E}$. In the right plot, we explore $R_K$ for a representative choice of masses that reproduces the $g-2$ anomaly: $M_{\cal F} = 1$~TeV, $M_{{\cal S}_E} = 0.9$~TeV and $M_{{\cal S}_D} = 1.5$~TeV. In both cases, $Y=1/2$. We maximise the Higgs contribution to the $\Delta a_{\mu}$ considering that the SM muon Yukawa coupling is radiatively generated following Eq~\eqref{eq:murad} (with $\Lambda = 5 M_{\cal F}$).}
\end{figure*}

It is useful to compare the composite scenario with its perturbative and elementary alter ego. The radiative explanation depends on three heavy particles masses, two scalars and one fermion. Furthermore, the ordinary Higgs Yukawa coupling to the new heavy fermions $\kappa$ yields the dominant contribution to the muon $g-2$ and therefore we further maximize it by requiring it to saturate the muon mass according to 
\begin{equation} \label{eq:murad}
m_\mu = N_{\rm TC}\frac{(y_{L} y_E^\dagger)_{\mu\mu} \kappa v_{\rm SM}}{(4 \pi)^2} \ln \frac{\Lambda^2}{M_{\cal F}^2} \ ,
\end{equation}
with $\Lambda$ a high energy cutoff interpreted as the scale at which the effective muon Yukawa vanishes. Without the Higgs corrections the model was investigated in \cite{Arnan:2016cpy} and because of this it cannot account for the observed $g-2$ of muon. 

We are now ready to illustrate the radiative results in Fig.~\ref{fig:loop}. After imposing the relation in Eq.~\eqref{eq:murad}, $\Delta a_\mu$  depends only on the masses of $\cal F$ and ${\cal S}_E$, and on $Y$. In the left panel, we show the allowed regions from the $g-2$ for $Y=1/2$. The green bands highlight an upper bound on the fermion mass at $1.5$~TeV, while low masses can also be reached at the price of heavier scalars. For comparison, we also show the region allowed by the BMW results.
To reproduce $R_K$ we must consider also the mass of the colour-triplet scalar ${\cal S}_D$: being charged under QCD, its mass is strongly constrained by the LHC searches, with precise bounds depending on the decays. We thus fix its mass to a benchmark value of $1.5$~TeV. In the right panel we show the dependence on the muon left-handed Yukawa and on the multiplicity $N_{\rm TC}$ for benchmark values of the masses that reproduce the muon $g-2$ anomaly. The general trend is that $R_K$ requires either large Yukawas or large field multiplicity, the latter constrained by the $B_s$ mixing. 
The main drawback of this scenario is that the Yukawas tend to develop a Landau pole at low scales, thus limiting the validity range of the model \cite{DAmico:2017mtc}.

To conclude, the anomalies associated to the muon physics can be successfully addressed within Technicolor-like models. The scale of new physics is around $2$~TeV implying that new massive states, such as the techni-rho, are  within  reach of the LHC direct searches \cite{Belyaev:2018jse}. The composite Goldstone Higgs scenario is disfavoured by the low composite scale and perturbative elementary extensions suffer from large muon left-handed Yukawa couplings and unnaturally large multiplicity of new fields. 

If the anomalies are confirmed by future experimental analyses, our results show that naturalness plays a fundamental role when constructing theories of nature, with impact in other realms from dark matter model building and searches to addressing the dark energy problem in cosmology. 

\section*{Acknowledgements}

GC and CC acknowledge partial support from the Labex LIO (Lyon Institute of Origins) under grant ANR-10-LABX-66  of the ANR (Agence Nationale pour la Recherche) and FRAMA (FR3127, F{\'e}d{\'e}ration de Recherche “Andr{\'e} Marie Amp{\`e}re”).
 
\appendix

\section{Appendix}

The loop functions used in Table~\ref{tab:cCH} have been extracted from \cite{Arnan:2016cpy,Calibbi:2018rzv} and adapted to our model. They are given by:
\begin{multline}
    F(x,y) = \frac{1}{(1-x)(1-y)} + \frac{x^2 \ln x}{(1-x)^2(x-y)} +\\  \frac{y^2 \ln y}{(1-y)^2(y-x)}\,,
\end{multline}
\begin{equation}
    F(x,x) = \frac{1-x^2+x\ln x}{(1-x)^3}\,,
\end{equation}
\begin{equation}
    F_{7}(y) = \frac{2+3y-6y^2+y^3+6y\ln y}{12(1-y)^4}\,,
\end{equation}
\begin{multline}
    \tilde{F}_{7}(y) = \frac{F_7 (y^{-1})}{y} \\ = \frac{1-6y+3y^2+2y^3+6y^2\ln y}{12(1-y)^4}\,,
\end{multline}
\begin{equation}
    F_{LR} (y) = \frac{1-y^2+2 y \ln y}{2(1-y)^3}\,,
\end{equation}
\begin{equation}
    G_{LR} (y) = \frac{1+4y+3y^2-2 y^2 \ln y}{2(1-y)^3}\,;
\end{equation}
where $x=M_{{\cal S}_D}^2/M_{\cal F}^2$ and $y=M_{{\cal S}_E}^2/M_{\cal F}^2$.

\bibliography{g2biblio}

\end{document}